\documentclass{emulateapj}
\RequirePackage{epstopdf}

\shorttitle{Sgr A East HII Regions}
\shortauthors{Lau et al. 2014}

\usepackage{graphicx}
\usepackage{amsmath}
\usepackage[section] {placeins}

\newcommand{\beq}{\begin{equation}}
\newcommand{\eeq}{\end{equation}}

\begin{document}

\submitted{Accepted to APJ August 23, 2014}

\title{Dusty Cradles in a Turbulent Nursery: The Sgr A East HII Region Complex at the Galactic Center}

\author{R. M. Lau\altaffilmark{1},
T. L. Herter\altaffilmark{1},
M. R. Morris\altaffilmark{2},
J. D. Adams\altaffilmark{1,3}
}

\altaffiltext{1}{Astronomy Department, 202 Space Sciences Building, Cornell University, Ithaca, NY 14853-6801, USA}
\altaffiltext{2}{Department of Physics and Astronomy, University of California, Los Angeles, 430 Portola Plaza, Los Angeles, CA 90095-1547, USA}
\altaffiltext{3}{SOFIA Science Center, Universities Space Research Association, NASA Ames Research Center, MS 232, Moffett Field, CA 94035, USA}

\begin{abstract}

We present imaging at 19, 25, 31, and 37 $\mu$m of the compact HII region complex G-0.02-0.07 located 6 pc in projection from the center of the Galaxy obtained with SOFIA using FORCAST. G-0.02-0.07 contains three compact HII regions (A, B, and C) and one ultra-compact HII region (D). Our observations reveal the presence of two faint, infrared sources located 23'' and 35'' to the east of region C (FIRS 1 and 2) and detect dust emission in two of the three ``ridges" of ionized gas west of region A. The 19/37 color temperature and 37 $\mu$m optical depth maps of regions A - C are used to characterize the dust energetics and morphology. Regions A and B exhibit average 19/37 color temperatures of $\sim 105$ K, and regions C and D exhibit color temperatures of $\sim115$ K and $\sim130$ K, respectively. Using the DustEM code we model the SEDs of regions A - D and FIRS 1, all of which require populations of very small, transiently heated grains and large, equilibrium-heated grains. We also require the presence of polycyclic aromatic hydrocarbons (PAHs) in regions A - C in order to fit the 3.6, 4.5, 5.8, and 8.0 $\mu$m fluxes observed by \emph{Spitzer/IRAC}. The location of the heating source for region A is determined by triangulation from distances and temperatures derived from DustEM models fit to SEDs of three different points around the region, and is found to be displaced to the northeast of the center of curvature near the color temperature peak. Based on total luminosity, expected 1.90 $\mu$m fluxes, and proximity to the mid-IR color temperature peaks we identify heating source candidates for regions A, B, and C. However, for region D, the observed fluxes at 1.87 and 1.90 $\mu$m of the previously proposed ionizing star are a factor of $\sim40$ times too bright to be the heating source and hence is likely just a star lying along the line of sight towards region D.

\end{abstract}

\maketitle

\section{Introduction}

The extreme environment in the Galactic center presents unique conditions for star formation given the hot, turbulent medium that is influenced by a strong tidal field, cloud-cloud collisions, stellar winds, and supernovae shocks (Morris \& Serabyn 1996). At least three young, clusters of massive stars exist within the inner 50 pc of the Galactic center (Nagata et al. 1995; Okuda et al. 1990; Krabbe et al. 1991), which provides strong evidence that the conditions in the region favor sudden events of massive star formation. Because molecular clouds in the central molecular zone require a high density in order to withstand the Galactic tidal force, the most extreme among them are prone to forming such massive clusters (Morris \& Serabyn 1996). Recently, however, numerous ($\sim$ 19) isolated Wolf-Rayet and O supergiants were found scattered throughout the inner 50 pc of the Galactic center with no association to the three massive clusters (Dong et al. 2011; Mauerhan et al. 2010; Dong et al. 2012). The existence of both the isolated stars and the massive, young clusters raises interesting questions about the conditions leading to massive star formation in the Galactic center region.

Located 6 pc in projection to the northeast of Sgr A* is a string of three compact HII regions and one ultra compact HII region (Ekers et al. 1983) known as G-0.02-0.07, or Sgr A East A-D. These HII regions appear coincident with the eastern edge of a prominent north-south ridge in the dense 50 km/s molecular cloud, M-0.02-0.07 (Goss et al. 1985; Serabyn et al. 1992), and exhibit identical radial velocities. Interestingly, this molecular cloud is apparently being impacted at the west by the Sgr A East supernova remnant (SNR); however, the star formation linked to the HII region complex, G-0.02-0.07, is completely unrelated to the SNR based on the dynamics of the SNR and the timescales required for the emergence of the HII regions (Serabyn et al. 1992). Despite not being a product of supernova-triggered star formation, G-0.02-0.07 is of great interest since it is a site of most recent confirmed star formation within $\sim10$ pc of the Galactic center.

The HII regions Sgr A East A-D are believed to each host a single late O-type star with stellar temperatures $\sim35000$ K (Serabyn et al. 1992) and exhibit ages of $\sim10^4-10^5$ years, with region D being the youngest given its nebular size. Studies of the extinction towards the HII regions suggest that regions A, B, and C lie on the surface of the 50 km/s molecular cloud while region D is more deeply embedded within the cloud (Serabyn et al. 1992; Mills et al. 2011). Observations of the dust emission from the HII regions can therefore aid in characterizing the properties of the heating sources and their interaction with the 50 km/s molecular cloud.

Recently, the HII regions have been studied by Yusef-Zadeh et al. (2010) and Mills et al. (2011) using observations of hot dust and ionized gas. Ground-based observations of the 12.8 $\mu$m [NeII] line revealed that the kinematics of regions A - C appear consistent with that of a bow shock produced by eastward moving O-stars through molecular gas, as opposed to a pressure-driven ``blister" resulting from a large density gradient from the molecular cloud to the surrounding medium (Yusef-Zadeh et al. 2010). Yusef-Zadeh et al. (2010) interpret the dual-lobed structure and the kinematics of region D as the result of a bipolar outflow collimated by an accretion disk. Based on Paschen-$\alpha$ (1.87 $\mu$m; Wang et al. 2010; Dong et al. 2011) observations of region D, Mills et al. (2011) agree with the bipolar outflow interpretation; however, Mills et al. (2011) find that the continuum source seen at 1.87 and 1.90 $\mu$m, which is slightly displaced from the 6 cm radio emission peak, is likely the central heating source of the region, as opposed to enhanced emission due to scattered light of a density clump as suggested by Yusef-Zadeh et al. (2010). Mills et al. (2011) generated a high-resolution extinction map of the HII regions that reveals the extinctions towards regions A - C are nearly identical ($A_V \sim45$) while $A_V\sim 70$ for region D.

In this paper we present 19.7, 25.2, 31.5, and 37.1 $\mu$m observations that trace the warm dust emission of the G-0.02-0.07 complex and address the presence of two faint infrared sources east of region C. These observations were taken by FORCAST aboard the Stratospheric Observatory for Infrared Astronomy (SOFIA). We discuss the morphology, energetics, and composition of the emitting dust in the G-0.02-0.07 complex to characterize the young stellar objects and explore the star formation environment in a dense molecular cloud at the Galactic center.

\section{FORCAST Observations}

Observations were made using FORCAST (Herter et al. 2012) on the 2.5-m telescope aboard SOFIA. FORCAST is a  $256 \times 256$ pixel dual-channel, wide-field mid-infrared camera sensitive from $5 - 40~\mu\mathrm{m}$ with a plate scale of $0.768''$ per pixel and field of view of $3.4'\,\times\,3.2'$.
The two channels consist of a short wavelength camera (SWC) operating at $5 - 25~\mu\mathrm{m}$ and a long wavelength
camera (LWC) operating at $28 - 40~\mu\mathrm{m}$. An internal dichroic beam-splitter enables simultaneous observation from both long and short wavelength cameras. A series of bandpass filters are used to image at selected wavelengths.

SOFIA/FORCAST observations of the Sgr A East HII complex were made on the OC1-B Flight 131 on September 17, 2013 (altitude $\sim$ 39,000 ft.) at 19.7, 25.2, 31.5, and 37.1 $\mu\mathrm{m}$. Measurements at 19.7 and 31.5 $\mu$m, as well as 25.2 and 31.5 $\mu$m, were observed simultaneously in dual-channel mode, while
the 37.1 $\mu$m observations were made in single-channel mode. Chopping and nodding were used to remove  the sky and telescope thermal backgrounds. An asymmetric chop pattern was used to place the source on the telescope
axis, which eliminates optical aberrations (coma) on the source. The chop throw was $7'$ at a frequency of $\sim 4$ Hz.
The off-source chop and nod fields (regions of low mid-infrared Galactic emission) were selected from the Midcourse Space Experiment (MSX) $21~\mu$m image of the Galactic center.

The source was dithered over the focal plane to allow removal of bad pixels and to mitigate response variations.
The on-source integration time at each dither position was $\sim 10$ sec, which includes inefficiencies from chopping. The images were reduced and combined at each wavelength according to the pipeline steps described in Herter et al. (2013). Total on-source integration time was $\sim100$  sec at 19.7 and 25.2 $\mu$m, $\sim200$ sec at 31.5 $\mu$m, and $\sim50$ sec at 37.1 $\mu$m. The quality of the images was consistent with near-diffraction-limited imaging at the longest wavelength; the full width at half maximum (FWHM) of the point spread function (PSF) was $3.2''$ at $19.7~\mu$m and $3.8''$ at 37.1 $\mu$m.

Calibration of the images was performed by observing standard stars and applying the resulting calibration factors as described in Herter et al. (2013). Color correction factors were negligible ($\lesssim5\%$) and were therefore not applied. The 1-$\sigma$ uncertainty in calibration due to photometric error, variation in water vapor overburden, and airmass is $\pm7\%$; however, due to flat field variations ($\sim15\%$), which we are unable to correct for, we conservatively adopt a 1-$\sigma$ uncertainty of  $\pm20\%$.

\section{Results and Analysis}

\begin{figure*}[c]
	\centerline{\includegraphics[scale=.6]{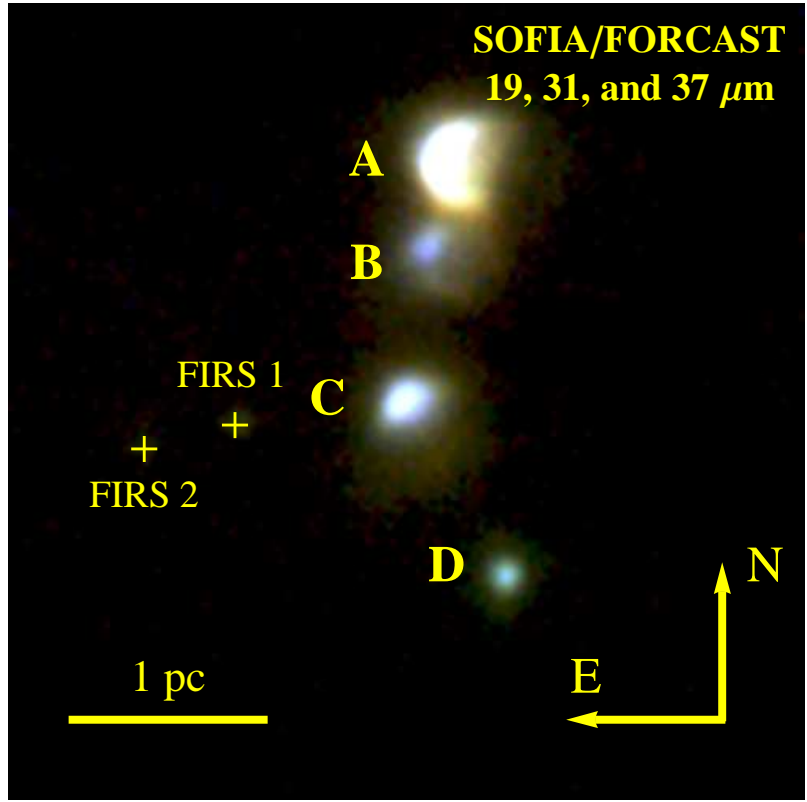}}
	\caption{19.7 (blue), 31.5 (green), and 37.1 (red) $\mu$m false color image of the Sgr A East HII Complex. The locations of the faint infrared sources (FIRS) 1 and 2 are given by the yellow crosses.}
	\label{fig:GCHIIColor}
\end{figure*}

A composite, false color image of the Sgr A East HII complex is shown in Fig.~\ref{fig:GCHIIColor} with the 19.7, 31.5, and 37.1 $\mu$m emission corresponding to the colors blue, green, and red, respectively. Fig.~\ref{fig:GCall}a - d shows the individual-band images of the complex at 19.7, 25.2, 31.5, and 37.1 $\mu$m.  The image reveals the warm dust emission from the four HII regions, Sgr A East A - D, two faint infrared sources (FIRS 1 and 2), and the presence of diffuse dust structures throughout the region.  Detections of regions A - D are made with a significance greater than 10-$\sigma$ above the background and FIRS 1 and 2 are detected at 6-$\sigma$ above the background. The 1-$\sigma$ levels at 19.7, 25.2, 31.5, and 37.1 $\mu$m are 0.013, 0.032, 0.015, and 0.031 Jy, respectively. Given the high signal-to-noise ratio of the HII regions, we perform a Richardson-Lucy deconvolution routine using a Gaussian with the horizontal (RA) full-width at half maximum (FWHM) of the point-source-like region D to characterize the point spread function (PSF). The deconvolution routine uses a uniform 3.0" FWHM Gaussian PSF at all wavelengths.

In this section we present and analyze the morphology, dust energetics, and heating source properties the HII regions and the faint IR sources. Their fluxes and properties are summarized in Tab.~\ref{tab:GCHIIFlux} and Tab.~\ref{tab:GCHIIProp}, respectively. We assume the Sgr A East HII complex to be located in the Galactic center at a distance of 8000 pc (Reid 1993).

\subsection{Interstellar Extinction}

Large column densities of dust and gas lead to extreme extinction along lines of sight towards the Galactic center ($A_V \sim 40$; Cardelli et al 1989). We adopt the extinction curve derived by Fritz et al. (2011) from hydrogen recombination line observations of the minispiral, the HII region in the inner 3 pc of the Galactic center, at 1 - 19 $\mu$m made by the Short Wave Spectrometer (SWS) on the Infrared Space Observatory (ISO) and the Spectrograph for Integral Field Observations in the Near Infrared (SINFONI) on the Very Large Telescope (VLT). We normalize the Fritz et al. (2011) extinction curve to the $A_{1.87}$ values derived by Mills et al. (2011) in their extinction study of the Sgr A East HII region complex from Paschen-$\alpha$ ($\lambda=1.87\,\mu$m) and 6 cm observations, yielding $A_{1.87}=3.7$ for regions A - C. For region D the 1.87 $\mu$m extinction is 2 magnitudes greater (Mills et al. 2011).

\subsection{Dust and Gas Morphology}

\subsubsection{Regions A - C}

\begin{figure*}[c]
	\centerline{\includegraphics[scale=.30]{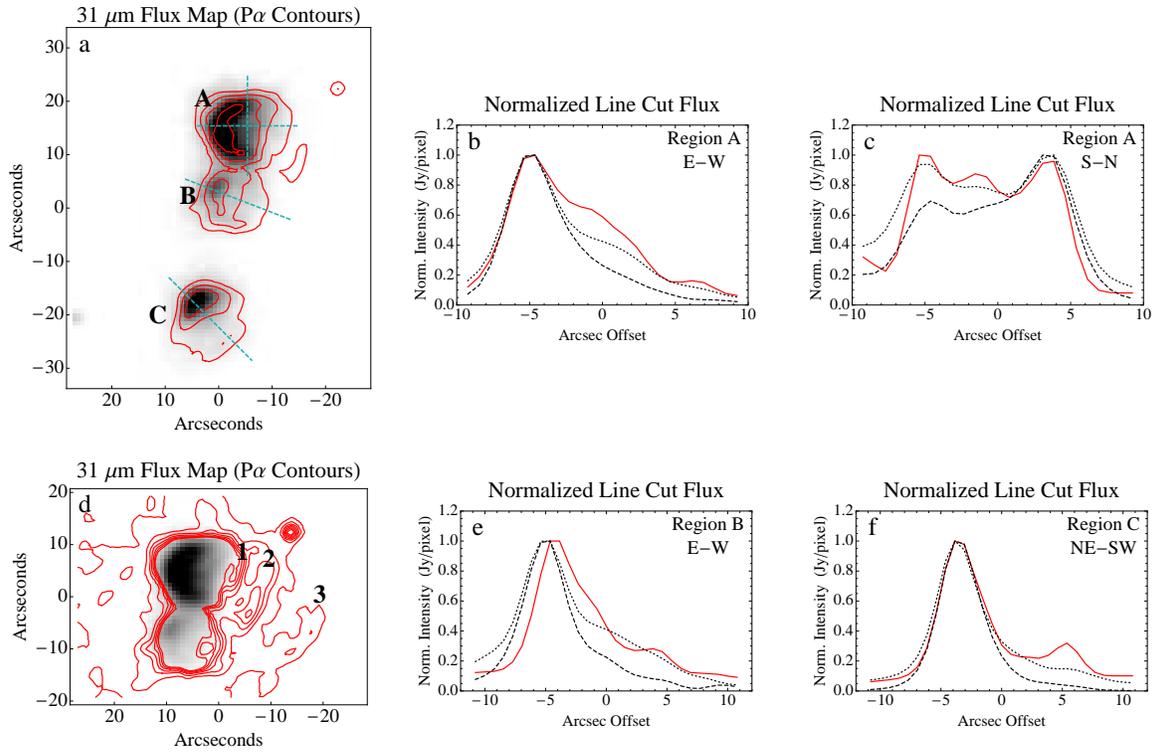}}
	\caption{(a) Paschen-$\alpha$ contours and line cuts overlaid on the 31.5 $\mu$m flux map of the Sgr A East HII Complex and normalized line cut flux plots through regions (b and c) A, (e) B, and (f) C of the Paschen-$\alpha$ (red solid), 19.7 $\mu$m (black dashed), and 31.7 $\mu$m (black dotted) emission. The Paschen-$\alpha$ emission has been convolved to match the beam size of the deconvolved 19.7 $\mu$m and 31.5 $\mu$m maps and the Paschen-$\alpha$ contour levels correspond to 0.25, 0.50, 1.0, and 2.0 mJy/pixel. (d) 31 $\mu$m flux map zoomed in on regions A and B overlaid with Paschen-$\alpha$ contours labeled with the locations of ``ridges" 1, 2, and 3 with levels corresponding to 0.10, 0.15, 0.20, 0.25, 0.30, 0.35, and 0.40 mJy/pixel.}
	\label{fig:GCHIICuts}
\end{figure*}

The 19.7 - 37.1 $\mu$m observations (Fig.~\ref{fig:GCall}a-d) resolve the extended nature of regions A - C. The dust emission at all the observed IR bands from region A exhibits a semi-circular, limb-brightened morphology with a diameter of $\sim0.45$ pc and is the brightest of all the sources in the complex. The ionized gas morphology traced by the Paschen-$\alpha$ emission agrees with the dust morphology at the eastern edge flux peak of the region, as shown in the contour overlay in Fig.~\ref{fig:GCHIICuts}a and the emission line cut profile in Fig.~\ref{fig:GCHIICuts}b; however, the south-north line cut profile (Fig.~\ref{fig:GCHIICuts}c) shows that the 19.7 $\mu$m emission at the southern edge decreases by $\sim40\%$ relative to the northern edge. This discrepancy between the 19.7 and 31.5 $\mu$m emission profiles strongly suggests a dust temperature gradient decreasing from north to south. Paschen-$\alpha$ emission extends slightly more than the dust into the central cavity.

Region B exhibits a slightly more complete circular morphology than region A, with a diameter of $\sim0.34$ pc. The dust emission at all observed IR bands peaks at an identical location on the northeastern edge of region B, but the Paschen-$\alpha$ flux peak is offset by several arcseconds to the west (Fig.~\ref{fig:GCHIICuts}e); however, the 31.5 $\mu$m emission peak agrees with the Paschen-$\alpha$ peak at the southern edge. Interestingly, of the three resolved HII regions, B is the only one that exhibits this displacement of the Paschen-$\alpha$ peak with respect to the dust emission peak. Similar to region A, both the 31.5 $\mu$m and Paschen-$\alpha$ emission extend further into the center of region B than the 19.7 $\mu$m emission. 

The dust in region C exhibits an elongated morphology along the southeast and northwest that appears more extended and concave with increasing wavelength, which is consistent with the concave morphology of the ionized gas. Although the effect is not as pronounced as regions A and B, the 31.5 $\mu$m and Paschen-$\alpha$ emission still appears to extend slightly further into the center of curvature of region C than the 19.7 $\mu$m emission. A linear filament located $\sim9.5''$ to the southwest of the dust emission peak of the region is apparent in the Paschen-$\alpha$ emission (Fig.~\ref{fig:GCHIIStars}) and is barely detected in the 31.5 and 37.1 $\mu$m maps. This southwestern filament appears elongated parallel to the main, bright segment of C. 

\subsubsection{Region D and The Faint IR Sources}

Region D is the most compact and southern source in the HII complex. The dust in the region exhibits a slight north-south elongation at 31.5 and 37.1 $\mu$m: the vertical FWHM is $\sim15\%$ larger than the horizontal FWHM at both wavelengths. At 19.7 and 25.2 $\mu$m the vertical FWHM is only $\sim5-10\%$  larger than the horizontal FWHM. This elongation observed from the dust emission is consistent with the north-south extension of the ionized gas morphology seen in Paschen-$\alpha$ emission.

\begin{figure*}[c]
	\centerline{\includegraphics[scale=.25]{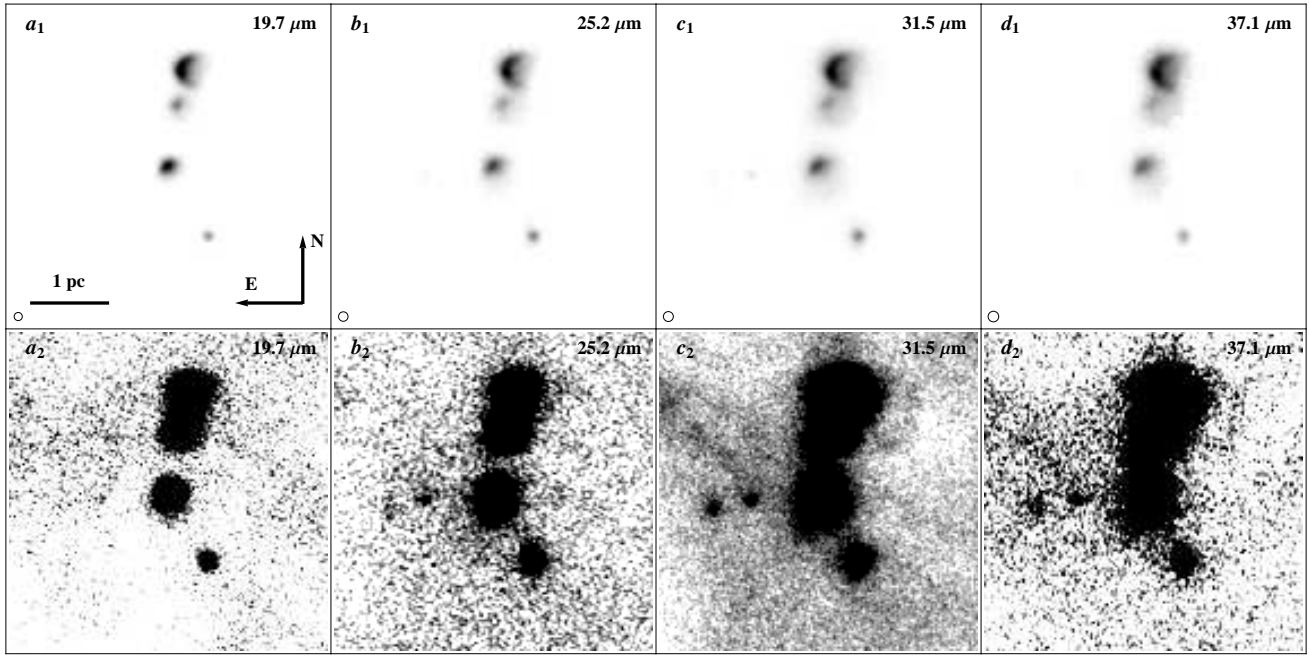}}
	\caption{Observed (a) 19.7, (b) 25.2, (c) 31.5, and (d) 37.1 $\mu$m images of the Sgr A East HII Complex at two different stretches to show bright (top panels) and faint (lower panels) emission. The approximate beamsizes are shown in the lower left corner in each image.}
	\label{fig:GCall}
\end{figure*}

Two faint IR sources (FIRS), designated 1 and 2, are detected 23'' and 35'' to the east of region C (Fig.~\ref{fig:GCall}). FIRS 1 is unresolved and has a $\sim15\%$ greater far-IR flux (19 - 37 $\mu$m; Tab.~\ref{tab:GCHIIFlux}) than FIRS 2 which appears partially resolved, with a FWHM of $\sim5.0''$ at 31.5 $\mu$m. Unlike the four HII regions, there is no detectable ionized gas counterpart in Paschen-$\alpha$ or 6 cm emission (Yusef-Zadeh et al. 1987) for FIRS 1, which suggests the heating source has a relatively cool effective temperature. FIRS 2, however, has a dim, shell-like ionized gas counterpart observed at 6 cm (Zhao, Morris, \& Goss 2014) with an angular diameter of $\sim5''$, consistent with the 31.5 $\mu$m FWHM. Emission from both faint IR sources is evident in the mid-IR at 5.8, and 8.0 $\mu$m by Spitzer/IRAC (Stolovy et al. 2006), and only FIRS 1 is observed at 4.5 $\mu$m. FIRS 2 is partially resolved At 5.8 and 8.0 $\mu$m with a FWHM of $\sim5.0''$, consistent with the FWHM observed at 31.5 $\mu$m. The Spitzer/IRAC fluxes from FIRS 1 are $\sim5-8$ times greater than that of FIRS 2. FIRS 1 appears coincident with the southern edge of a long linear filament observed at 31.5 and 37.1 $\mu$m southeast of regions A - C (Fig.~\ref{fig:GCall}c and d). Ionized gas emission is also detected from this filament in Paschen-$\alpha$, and it has been suggested that the filament traces the surface of the 50 km/s molecular cloud (Mills et al. 2011).

The mid-IR flux and position of FIRS 1 are arising from consistent with arising from the previously identified OH/IR star and SiO maser, SLV2002 SiO 359.977-0.087 (Sjouwerman et al. 2002), as well as SSTGC 564417, a source identified in the YSO search performed by An et al. (2011) with observations from Spitzer/IRAC. The position of FIRS 2 is consistent with the X-ray point source CXOUGC J174554.4-285809 (Muno et al. 2009).

\subsubsection{Western ``Ridges"}

Three faint ionized curved ``ridges" are located approximately 7'' (ridge 1), 14'' (ridge 2), and 26'' (ridge 3) to the west of region A (Mills et al. 2011), the most extended of which can be seen in Fig.~\ref{fig:GCall}c and d. Only the two furthest ridges from region A are resolved and detected at 31.5 and 37.1 $\mu$m at a flux level greater than 4-$\sigma$ and 2-$\sigma$ for ridges 2 and 3, respectively. Ridge 1 is confused with the extended dust emission from region A. The dust emission from ridges 2 and 3 extends to angular sizes of $\sim17''$ and $\sim31''$, respectively, and is roughly cospatial with their apparent Paschen-$\alpha$ counterparts; however, the dust emission from ridge 3 extends further north and arcs back towards region A, whereas its Paschen-$\alpha$ counterpart appears to end at the same declination as the southern edge of region A. 

The Paschen-$\alpha$ and 6 cm emission measurements of ridges 2 and 3 indicate that there is additional extinction along lines of sight towards ridge 3. Extinction estimates towards ridge 2 are found to be nearly identical to that of regions A - C, which appears consistent with the extinction map produced by Mills et al. (2011). The 1.87 $\mu$m extinction increases by $\Delta \tau_{P_\alpha}\lesssim 0.8$ magnitudes between ridges 2 and 3. Comparison of the 1.87 $\mu$m extinction between ridges 1 and 2 reveals that ridge 1 has $\sim 0.6\pm0.2$ magnitudes less extinction. It is important to note that the non-thermal radio emission from the supernova remnant Sgr A East may be contaminating the thermal emission from ridge 3. A subtraction of the non-thermal emission was attempted when calculating the extinction; however, there are large uncertainties since the emission is non-uniform and the extinction towards ridge 3 should therefore be treated as an upper limit. The projected locations, extinction, dust properties of the ridges are provided in Tab.~\ref{tab:GCHIIA}.

\subsection{Color Temperature and Optical Depth}

The color temperature maps are derived from the ratio of the deconvolved 19.7 and 37.1 $\mu$m flux maps assuming the emission is optically thin and takes the form of $F_\nu \propto B_\nu(T_d)\,\nu^\beta$, where a value of 2 for $\beta$ is adopted. With the dust temperature and deconvolved 37.1 $\mu$m flux maps, the 37.1 $\mu$m optical depth over the pixel solid angle, $\Omega_p$, is simply:

\beq
\tau_{37.1}= \frac{F_{37.1}}{\Omega_p\,B_\nu (T_d)}.
\label{eq:OD}
\eeq

\begin{figure*}[h]
	\centerline{\includegraphics[scale=.4]{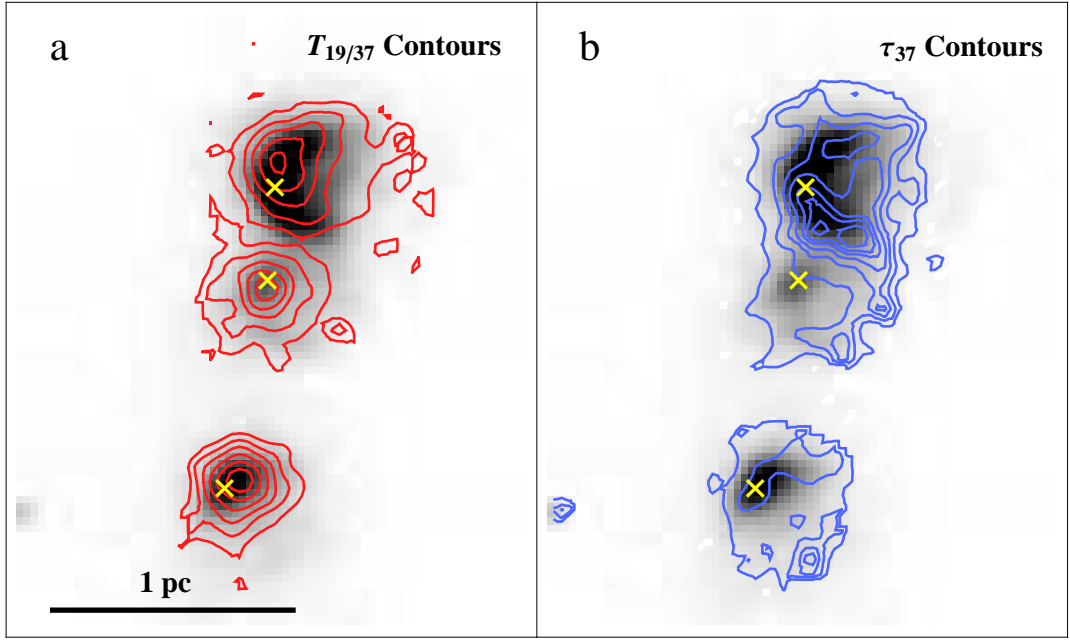}}
	\caption{31 $\mu$m flux map of the Sgr A East HII Complex overlaid with contours of the (a) 19/37 dust temperature and (b) 37 $\mu$m optical depth. The location of the 31 $\mu$m flux peak in each region is indicated by the yellow cross. The levels of the temperature contours correspond to 90, 100, 110, 120, 130, and 140 K. The levels of the 37 $\mu$m optical depth contours correspond to 0.001, 0.0025, 0.004, 0.0055, 0.0075, 0.009, and  0.011.}
	\label{fig:GCTempandOD}
\end{figure*}

The color temperature and column density contours of the complex are overlaid on the 31.5 $\mu$m flux map in Fig.~\ref{fig:GCTempandOD}.

The color temperature of region A peaks at a value of 135 K at the location $\sim4''$ north of the dust emission peak at the eastern edge, and exhibits an average temperature of $107\pm7$ K. Westward of the peak, the temperature of region A decreases approximately linearly to a value of $\sim90$ K at a distance of 12'' from the peak. The symmetry of the color temperature map is significantly different from the 31.5 $\mu$m emission map: the dust emitting at the north is $\sim15$ K hotter than the dust at the south while the flux from those regions are similar within $\sim15\%$.

The 37.1 $\mu$m optical depth of region A peaks at the southern edge of the dust emission at a value of $\sim0.011$. Northwards and clockwise along the emitting dust, the optical depth decreases to a minimum of $\sim0.005$ at approximately the 10 o'clock position and then increases to $\sim0.006$ at 11 and 12 o'clock positions along the northern edge. Despite the symmetry of the dust emission at 31.5 $\mu$m, a majority of the dust is located at the southern edge of the region. Similar to the dust emission intensity, the optical depth sharply decreases to the east of the region. To the west of the center of curvature there is a slight increase in the optical depth before falling off sharply.

Region B exhibits a color temperature peak of 135 K located at the northeast dust emission edge of the region and within $\sim2''$ of the dust emission intensity peak. The average temperature is similar to that of region A at a value of $106\pm6$ K. The temperature decreases symmetrically in all directions to a value of $\sim85$ K at a location 7'' from the peak. Interestingly, the 37.1 $\mu$m optical depth of region B appears almost perfectly anti-correlated with both the dust and ionized gas emission. The dust and ionized gas emission peak along the eastern edge of the region; however, the optical depth is lowest at the east and increases towards the west along the rim of the shell to values $\sim3$ times that of the east.

The dust temperature of region C peaks at a value of 145 K and is consistent with the location of the Paschen-$\alpha$ emission peak, but is slightly offset by $\sim3''$ to the northwest of the dust emission peak. The region exhibits an average temperature of $115\pm10$ K and decreases to $\sim75$ K at the linear filament to the southwest (Fig.~\ref{fig:GCHIICuts}f and Fig.~\ref{fig:GCHIIStars}c). The 37.1 $\mu$m optical depth contours trace the concave morphology of the Paschen-$\alpha$ and dust emission. At the southwest filament the optical depth is approximately twice that of the brighter, main region in C, which is similar to what is observed in region B.

After correcting for the additional extinction ($\Delta A_{P\alpha}\sim2$), region D exhibits a 19/37 color temperature of 130 K. Assuming the extinction towards FIRS 1 and 2 is identical to that of regions A - C, they would both exhibit color temperatures of $\sim90$ K. However, dust models fit to the SEDs of FIRS 1 suggest that there is much more intervening dust than along the lines of sight towards Region A - C ($\Delta A_{P\alpha}\sim7$; Sec.~3.4.4). Dereddening the 19 and 37 $\mu$m flux of FIRS 1 with $\Delta A_{P\alpha}\sim7$ provides a 19/37 color temperature of 173 K.

The 31/37 color temperatures of ridges 2 and 3, located to the west of region A, decrease with projected distance from region A: ridges 2 and 3 exhibit temperatures of 63 and 40 K, respectively. The 37.1 $\mu$m optical depths increase with projected distance from region A: ridges 2 and 3 are $\sim0.01$ and $\gtrsim0.04$, respectively. Since 37.1 $\mu$m is far on the Wien side of the Planck function at a temperature of $T=40$ K and the uncertainty in the 31/37 color temperature is $\pm4$ K, the 37.1 $\mu$m optical depth for ridge 3 is not well constrained; however, even at the lower limit the 37.1 $\mu$m optical depth of ridge 3 is still a factor of 4 times greater than that of ridge 2.

\subsection{SEDs and Dust Modeling}

To better understand the nature of the grains and their thermal behavior, we construct physical models of the grain energy balance and subsequent emission spectrum with the aim of reproducing the observed SEDs. This is done for both spatially integrated SEDs for all the sources (Fig.~\ref{fig:GCHIISEDs}) and for selected regions of source A (Fig.~\ref{fig:GCASEDs}). A detailed spatial model for source A is constructed based on the dust properties of the regional models (Fig.~\ref{fig:GCAPlots}). Details of the models and the resulting constraint on dust properties are given in this section.

\subsubsection{Evidence of Very Small Grains}

\begin{figure*}[h]
	\centerline{\includegraphics[scale=.27]{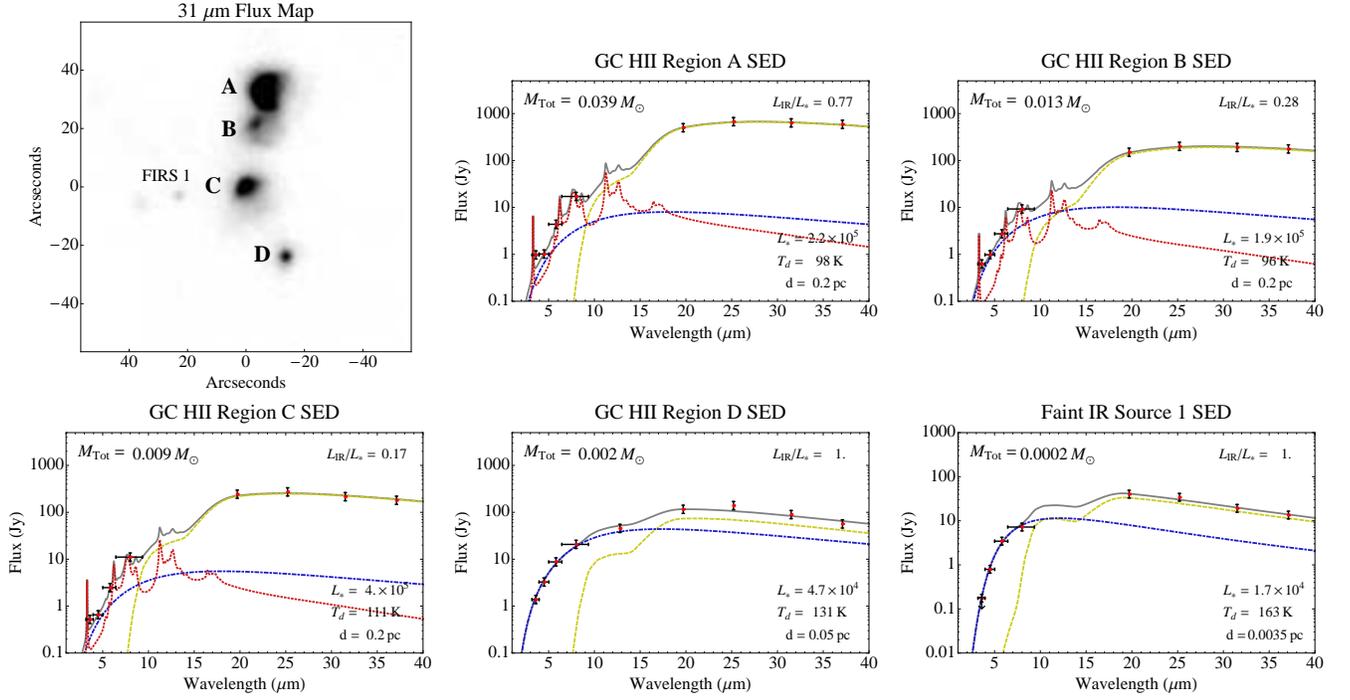}}
	\caption{31 $\mu$m flux map of the Sgr A East HII Complex and DustEM fits (solid grey line) to the dereddened SEDs of regions A - D and the FIRS 1. The DustEM models of regions A - D assume the dust is composed of LGs ($a_{LG}:100-1000\,\AA$;), VSGs ($a_{VSG}:6-25\,\AA$), and PAHs ($a_{PAH}:6-10\,\AA$). DustEM models of FIRS 1 assume the same dust composition as the HII region models except the maximum grain size cutoff of the VSGs is 100 $\AA$. The dotted red lines, dot-dashed blue lines, and dashed yellow lines correspond to the PAH, VSG, and LG emission components, respectively. The total dust mass, $M_\mathrm{Tot}$, and the infrared to total heating source luminosity ratio, $L_{IR}/L_*$, are given at the top of each plot. The heating source luminosity, $L_*$, the volume-average dust temperature, $T_d$, and the distance between the heating source and the dust, $d$, are given at the bottom right of each plot.}
	\label{fig:GCHIISEDs}
\end{figure*}

The inclusion of the mid-IR observations of the HII regions and FIRS 1 taken by Spitzer/IRAC (Stolovy et al. 2006) in their SEDs (Fig.~\ref{fig:GCHIISEDs}) reveals that dust models composed of a distribution of strictly large grains ($a=100 - 1000 \,\AA$; dotted yellow line in Fig.~\ref{fig:GCHIISEDs}b - f), which agree with the observed flux at 19 - 37 $\mu$m, fail to fit the observed flux at 3.6 - 8 $\mu$m by more than an order of magnitude. The enhanced emission at the IRAC wavelengths is attributed to the presence of very small ($a<100\, \AA$), transiently heated grains in the region. Very small grains (VSGs) such as Polycyclic Aromatic Hydrocarbons (PAHs; $a\sim4 - 12\,\AA$) can reach temperatures much greater than equilibrium-heated large grains since they have small heat capacities that result in large temperature spikes after absorbing single photons. The emissivity of small grains can be characterized as

\beq
j_\nu=\int \mathrm{d}a \frac{\mathrm{d}n}{\mathrm{d}a} \int \mathrm{d}T \left(\frac{\mathrm{d}P}{\mathrm{d}T}\right)_a \sigma_\mathrm{abs}(\nu\mathrm{, }\,a)B_\nu(T),
\eeq

where $n$ is the dust density and $ \frac{\mathrm{d}P}{\mathrm{d}T}$ is a probability distribution function with $P(T)$ being the probability that a grain will have a temperature less than or equal to $T$. For large grains in radiative equilibrium the probability distribution function is simply a delta function at the equilibrium temperature. The steady state probability distribution function for small grains, which is much broader than the large grain function, can be solved for directly (Guhathakurta \& Draine 1989; Draine \& Li 2001). 

\subsubsection{DustEM SED Models}

The DustEM code (Compi{\`e}gne et al. 2011) is utilized to model the SEDs of the HII regions and the faint IR sources. DustEM is capable of modeling the emission from stochastically heated VSGs using the formalism of Desert et al. (1986) to derive the temperature probability function of the grains. In the models, it is assumed that there are three populations of dust in the region: large silicate grains (LGs), VSGs composed of carbonaceous material, and PAHs. For the models of regions A - D, the LGs, VSGs, and PAHs have sizes ranging from 100 - 1000 $\AA$, 6 - 25 $\AA$, and 6 - 10 $\AA$ respectively. VSG size distributions with larger upper grain size cutoffs such as $a_\mathrm{max}=100$ $\mu$m result in very poor or physically unreasonable fits to the SEDs of regions B and D. Regions A and C models with an upper VSG size cutoff of 100 $\mu$m do not predict the same relative dust mass abundances as the $a_\mathrm{max}=25$ $\mu$m models; however, they predict dust temperatures, stellar luminosities, and total dust masses identical to the $a_\mathrm{max}=25$ $\mu$m models. The dust model for FIRS 1 is identical to the previous models except the VSGs range from 6 - 100 $\AA$ in size since the 6 - 25 $\AA$ VSG model over-predicts the emission at 3.6 $\mu$m. The minimum grain sizes for the VSGs and PAHs are determined such that their maximum temperature reached is less than their sublimation temperatures, which is $\sim1800$ K for the carbonaceous VSGs and $\sim1500$ K for PAHs (Draine 2011). 

 A grain size distribution power-law index of $\alpha=-3$, which is similar to the index adopted for the interstellar dust model from Desert et al. (1990), is assumed for all the grain populations. A standard MRN distribution grain size power law index of $\alpha =-3.5$ (Mathis, Rumpl, \& Nordsieck 1977) provides similar results to $\alpha=-3$. 
 
Castelli \& Kurucz (2004) stellar atmospheres are utilized to model the radiation field heating the dust. Initially, we assume the heating sources have stellar atmospheres consistent with the late-type O-stars predicted by Mills et al. (2011) from their estimates of ionizing flux from the regions: an O7 model for the region A heating source, O8.5 for regions B and C, and O9 for region D. However, for regions A - C, we find that the main-sequence luminosity of these model stars ($L_*\sim6 - 10\times10^4\,L_\odot$) under-predicts the dust temperatures as compared to the observations (see Sec.~4.5). The stellar luminosity is therefore treated as a free parameter in our models for regions A - C, which exhibit an ``open" shell-like morphology (Fig.~\ref{fig:GCHIICuts}). After an initial model fit, the newly derived stellar luminosities were used to reclassify the spectral type of the heating source based on the stellar classification of Martins et al. (2005), which was also adopted by Mills et al. (2011). The stellar atmospheres of the heating sources assumed in the final best-fit model are an O6 atmosphere for regions A and B, an O5 atmosphere for region C, and an O9 atmosphere for region D. Since there is no radio or Paschen-$\alpha$ counterpart to FIRS 1, which indicates the absence of ionized gas, a $T_*=7000$ K blackbody stellar atmosphere is adopted for the FIRS 1 heating source. 

For regions A - C, the DustEM models are fit to the SEDs with the stellar luminosity, $L_*$, and the grain mass abundances $Y_\mathrm{LG}$, $Y_\mathrm{VSG}$, and $Y_\mathrm{PAH}$ as free parameters, and the dust is assumed to be confined to a shell located 0.2 pc from the heating source. The free parameters for the region D models are similar to those of the regions A - C models, where the differences are the inclusion of the heating source distance as a free parameter and fixing the $L_\mathrm{IR}/L_*$ ratio to a value of 1 since region D is likely embedded. The FIRS 1 models include the same free parameters as the region D models as well as an additional parameter, $\Delta A_{P\alpha}$, the local extinction relative to the extinction along the lines of sight towards regions A - C. Similar to the region D models, the IR to stellar luminosity ratio is fixed at 1 for FIRS 1. Models fit to the FIRS 2 SED are very poorly constrained since the luminosity of the heating source and its distance to the dust in FIRS 2 are not known; therefore, DustEM analysis of FIRS 2 is omitted in this paper.

The DustEM model fits to the SEDs of the HII regions and FIRS 1 are shown in Fig.~\ref{fig:GCHIISEDs} and the model parameters are given in Tab.~\ref{tab:GCHIIProp}. The dust temperature provided, $T_d$, is the volume-average dust temperature,

\beq
T_d=\frac{\int \mathrm{d}a \frac{\mathrm{d}n}{\mathrm{d}a} \,T(a)\, m(a)}{\int \mathrm{d}a \frac{\mathrm{d}n}{\mathrm{d}a}\,m(a)}.
\label{eq:Td}
\eeq

In the models $dn/da \propto a^{-3}$ and $m(a) = 4/3 \pi a^3 \rho_b$, where $\rho_b$ is the bulk density of the dust grains. Eq.~\ref{eq:Td} can then be reduced to

\beq
T_d=\frac{\int \mathrm{d}a \,T(a)}{\int \mathrm{d}a}.
\label{eq:Td2}
\eeq

The resulting integrated IR-to-stellar luminosity ratios derived for regions A - C ($L_\mathrm{IR}/L_*$), which can be related to the fractional coverage of the dust around the heating source, agree with the apparent dust morphology of each region: region A appears as an almost complete spherical shell, which suggests a high fractional coverage ($L_\mathrm{IR}/L_*\sim0.77$) while C appears as an arc with a large opening angle which suggests a lower fractional coverage ($L_\mathrm{IR}/L_*\sim0.17$). The stellar luminosities fit to the heating sources of regions A - C predict ionizing fluxes $\sim1.5 - 7$ $\times$ greater than the ionizing fluxes determined by Mills et al. (2011), which can be attributed to the morphology and fractional coverage of the regions. It follows from this reasoning that the ionizing flux we derive for region D is consistent with the estimate from Mills et al. (2011) since it is a fully embedded source.

\subsubsection{Presence of PAHs}

The presence of PAHs is required for the DustEM model fits to the SEDs of regions A - C due to the enhanced 3.6 $\mu$m emission. The Spitzer/IRAC bandwidth of the 3.6 $\mu$m filter includes the 3.3 $\mu$m PAH feature. Without PAHs in the dust models, the VSG emission alone would fall short by greater than an order of magnitude from the observed 3.6 $\mu$m emission. As can be seen in Fig.~\ref{fig:GCHIISEDs}, the fits to the region D and FIRS 1 SEDs do not require any PAHs; however, the 1-$\sigma$ upper limit of $\frac{Y_\mathrm{PAH}}{Y_\mathrm{VSG}}$ for region D is as high as $\sim0.3$, which is comparable to $\frac{Y_\mathrm{PAH}}{Y_\mathrm{VSG}}$ for regions A - C. The disparity of PAHs is more constrained in the FIRS 1 models, where the 1-$\sigma$ upper limit of $\frac{Y_\mathrm{PAH}}{Y_\mathrm{VSG}}$ is $\sim0.02$.

\subsubsection{Relative Extinction Towards Faint IR Source 1}

Fits to the FIRS 1 SED require an additional extinction relative to the extinction along the line of sight towards regions A - C, where $A_{P\alpha}=3.7$ (Mills et al. 2011). In the FIRS 1 models, the additional extinction is taken to be at the Paschen-$\alpha$ wavelength (1.87 $\mu$m) and is designated as $\Delta A_{P\alpha}$. The best-fit models for the SED require $\Delta A_{P\alpha}=7\pm2$. Neither the upper nor the lower 1-$\sigma$ models require the presence of PAHs. The estimated luminosity of the heating source is $1.7_{-0.6}^{+1.2}\times10^4$ $L_\odot$.

\subsubsection{Region A SEDs and Heating Source Location}

We now determine the location of the heating source in region A. Assuming the composition of the emitting dust does not change appreciably across the source, we can pick three points within the nebula and triangulate the position of the heating source. The distance, $d$, from the heating source determines its temperature from energy balance, which for spherical, large grains is:

\beq
\pi  a^2Q_*(a)\, \frac{L_*}{4\pi d^2}=4 \pi  a^2\,Q_{\mathrm{dust}}(T_d,a)\,\sigma _{\mathrm{SB}}\,T_d{}^4,
\label{eq:Deq}
\eeq

where $Q_\mathrm{dust}$ is the dust emission efficiency averaged over the Planck function of dust grains of size $a$ and temperature $T_d$, $Q_*$ is the dust absorption efficiency averaged over the incident radiation field, and $L_*$ is the total integrated stellar luminosity. Taking silicate grains with a 0.01 $\mu$m radius heated radiatively by an O6 star, the radiation-field-averaged absorption efficiency, $Q_*$, is $\sim0.5$ and the dust-averaged emission efficiency can be expressed as $Q_\mathrm{dust}(a)=1.3\times10^{-7}\frac{a}{0.01\,\mu\mathrm{m}}$ (Draine 2011). 

\begin{figure*}[h]
	\centerline{\includegraphics[scale=.30]{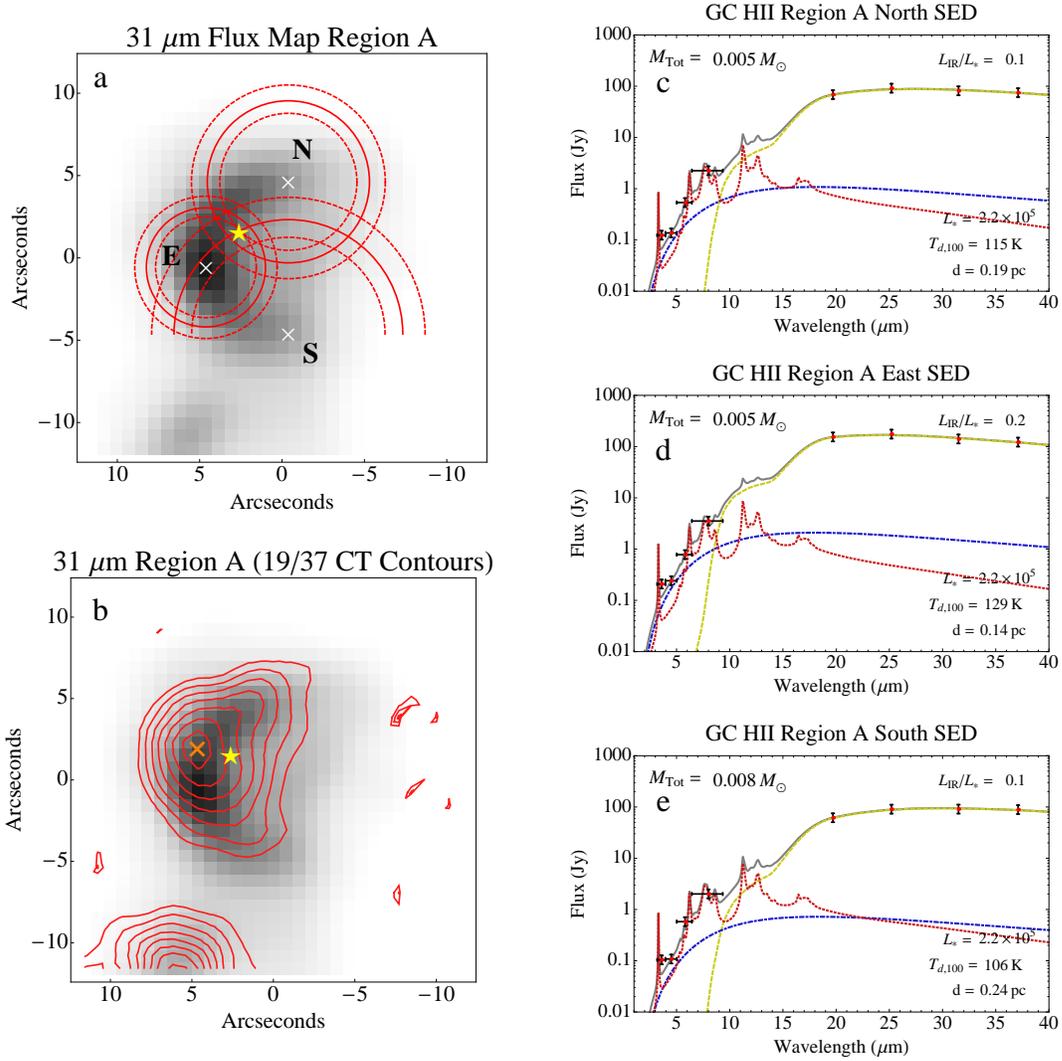}}
	\caption{(a) 31 $\mu$m flux map of region A overlaid with annuli representing the possible locations of the heating source centered on the north, east, and south positions, which are marked by white $\times$'s. The outer and inner dashed rings represent the 1-$\sigma$ ($\sim20\%$) uncertainty in the inferred heating source luminosity. The star indicates the implied location of the heating source, the intersection of the three annuli. (b) 31 $\mu$m flux map of region A (greyscale) overlaid with the location of the peak 19/37 color temperature, which is shown as the orange ``$\times$", and contours corresponding to 100, 105, 110, 115, 120, 125, and 130 K. (c) - (e) DustEM model fits (solid grey line) to SEDs at the north, east, and south positions of region A . $T_{d,100}$ is the temperature of the 100 $\AA$-sized grains determined by the DustEM models. The dotted red lines, dot-dashed blue lines, and dashed yellow lines correspond to the PAH, VSG, and LG emission components, respectively.}
	\label{fig:GCASEDs}
\end{figure*}

The location of the heating source is given by the star in Fig.~\ref{fig:GCASEDs}a and b which is at the intersection of the three annuli that represent the radius from the three regions at which the heating source must be located at to be consistent with the model dust temperature of the 100 $\AA$-sized grains, $T_d$. The outer and inner dashed rings show the radii at $1-\sigma$ deviations of the stellar luminosity. Since the highest dust temperatures are exhibited at the north and east of region A, the heating source is inferred to be displaced $\sim3''$ to the north east of the center of curvature, which is consistent with the location of the peak color temperature given by the orange `$\times$'.

DustEM models were fit to the SEDs from the north, east, and south sections of region A (Fig.~\ref{fig:GCASEDs}). Grain properties, heating source luminosity, and effective temperature were fixed to the values determined for the region A model. The distance from the dust to the heating source is as determined above. The free parameters are the relative abundances of the three grain populations. Local mechanisms that increase with density such as dust backscattering and trapped Lyman-alpha are minor contributors to the heating and thus are not responsible for the dust temperature asymmetry. The dust mass determined at the south is $\sim2$ times greater than the mass at the north and east, which agrees with the positive north-south gradient in the 37.1 $\mu$m optical depth contours (Fig.~\ref{fig:GCTempandOD}b). 

\subsubsection{Region A Dust Emission Model}

In order to understand the structure and heating of the dust in region A, we generated a dust emission model assuming the dust is arranged in a hemispherical shell and is heated by a single O6 star. Three different gas densities are used to model the emission: $n_S$ is the density of the southern half of the shell, $n_C$ the central quarter, and $n_N$ the northern quarter. The densities are related to each other as

\beq
n_S = 2 n_C,\,
n_N = 1.5 n_C,
\eeq

which is consistent with the relative column densities in the 37.1 $\mu$m optical depth map. The other free parameters are the shell thickness, $\Delta t$, and inner shell radius, $r_0$. A fixed, uniform grain size distribution of $0.01$ $\mu$m-sized silicates with a gas-to-dust mass ratio of 100 and a flat density profile radially through the shell is assumed for the model. The location of the heating source is as determined in the previous section: $3''$ to the east and $1''$ to the north of the center of curvature. Parameters for the emission model are given in Table~\ref{tab:GCHIIAMod}.

The 31 $\mu$m emission model, which has been convolved to the resolution of the observed 31 $\mu$m map, is shown in Fig.~\ref{fig:GCAPlots}a and is overlaid with the model 19/31 color temperature contours. Both the color temperature contours and the 31 $\mu$m dust intensity of the model agree very well with the observed temperature contours and dust intensity (Fig.~\ref{fig:GCAPlots}b), especially the displacement of the intensity peak from the temperature peak. Fig.~\ref{fig:GCAPlots}c shows the close agreement of the vertical (red) and horizontal (blue) line cuts through both the model (solid) and observed (dotted) 31 $\mu$m maps.

\subsubsection{Gas-to-Dust Mass Ratio}

The dust mass of the HII regions and FIRS 1 can be derived from the DustEM models based on the number of LGs, VSGs, and PAHs required to fit the observed SEDs, although in general the dust mass is dominated by the LGs. Dust mass estimates for the HII regions are consistent with the mass derived from the observed integrated column densities of each region assuming a uniform distribution of 0.01 $\mu$m-sized grains. HII masses estimated from Paschen-$\alpha$ and 6 cm observations (Mills et al. 2011) give gas-to-dust mass ratios of 56, 60, 45, and 65 for regions A, B, C, and D, respectively. This estimate will be a lower limit since dust will be heated at the molecular cloud interface outside the HII region.

\section{Discussion}

\subsection{Dust Heating by Trapped Lyman-$\alpha$}

Lyman-$\alpha$ photons within dusty HII regions are rapidly absorbed and reemitted by hydrogen atoms until they are eventually absorbed by the dust grains. This dust heating mechanism is therefore proportional to the free-free emission from the region. The contribution to the infrared ($\sim6-40$ $\mu$m) luminosity from the dust, $L_\alpha$, can be expressed as

\beq
L_\alpha=\frac{S_{\nu,\mathrm{obs}}}{j_{\nu,ff}(n_e,T_e)}n_e^2(\alpha_B - \alpha_{2s1})h \nu_{Ly\alpha}\, 4 \pi d^2,
\label{eq:Lyalpha}
\eeq

where $S_{\nu,\mathrm{obs}}$ is the observed radio continuum flux, $\alpha_B$ is the type-B recombination coefficient, and $\alpha_{2s1}$ is the recombination coefficient for the two-photon process ($\alpha_{2s1}<\alpha_B$; Osterbrock \& Ferland 2006). Given the electron temperatures ($\sim6000 - 7000$ K; Goss et al. 1985) and the 8.4 GHz flux ($\sim0.1 - 0.6$ mJy; Mills et al. 2011) of the regions, we find that the trapped Lyman-$\alpha$ heating contributes only $\sim10\%$ of the total IR luminosity derived from the DustEM models (Tab.~\ref{tab:GCHIIProp}). If the dust heating from trapped Lyman-$\alpha$ were more dominant, the dust temperature would trace the density distribution and there would be a strong correlation between the color temperature and the optical depth maps (Fig.~\ref{fig:GCTempandOD}); such a correlation is not observed since the dust heating is dominated by the stellar radiation.

\subsection{Region A: Bow Shock or Blister}

The difference in the pressure between the ionized gas and the surrounding medium as well as high velocity ($\sim2000$ km/s) stellar winds from the ionizing star can play important roles in driving the expansion of HII regions. By comparing the rate of expansion from both mechanisms, Shull (1980) found that stellar winds dominate the dynamics over gas pressure if the following condition is satisfied

\beq
\left(\frac{L_w}{10^{36}\,\mathrm{ergs}\,\mathrm{s}^{-1}}\right)>0.15\left(\frac{Q_0}{10^{48}\,\mathrm{s}^{-1}}\right)^{2/3}\left(\frac{n_0}{10^4\,\mathrm{cm}^{-3}}\right)^{-1/3},
\label{eq:wind}
\eeq

where $L_w$ is the mechanical power of the wind, $Q_0$ is the number of ionizing photons emitted per second, and $n_0$ is the density of the molecular cloud. For a star with a mass-loss rate of $10^{-6}\,M_\odot/yr$ and wind velocities of $\sim2000$ km/s, which are typical values for massive O-type stars, the mechanical power of the wind is $\sim10^{36}$ erg/s. Assuming values consistent with Mills et al. (2011) of $n_0\sim10^5$ $\mathrm{cm}^{-3}$ and $Q_0$ $\sim 10^{48}\,\mathrm{s}^{-1}$ for the HII regions, Eq.~\ref{eq:wind} is satisfied for all and their expansion is likely dominated by the stellar winds. This is consistent with the interpretation of the position-velocity data made by Yusef-Zadeh et al. (2010), where they conclude that the regions A - C are in fact bow shocks driven by stellar winds and the relative motion of the stars through the molecular cloud as opposed to ``blisters" resulting from pressure driven flows at the surface of the molecular cloud.

The agreement between the dust emission model, location of the 19/37 color temperature peak, and DustEM model fits to the SEDs around the region strongly suggest that the heating source is displaced to the northeast from the center of curvature. This northeast offset of the star is likely due to its motion relative to the surrounding medium. The combined effects of the northeast motion of the star and its stellar winds are producing the bow shock that characterizes region A. This interpretation agrees with the steep radial velocity gradient at the east (Yusef-Zadeh et al. 2010), whereas a ``blister" would produce the opposite signature, with a steep velocity gradient at the west, which is the direction of the decreasing pressure gradient from the cloud to the ambient medium. By analyzing the kinematics of region A, Yusef-Zadeh et al. (2010) infer that the ionizing star is moving at a velocity of $\sim30$ km/s towards the east and projected in our direction by an angle of $30^\circ$ and is likely driving a bow shock into the 50 km/s molecular cloud.

In a bow shock, the stagnation radius--the distance from the star to where the momentum flux of the wind balances the ram pressure of the surrounding medium--can be expressed as 

\begin{multline}
r_s \sim  0.04 \left(\frac{\dot{M}}{10^{-6}\,M_\odot \, \mathrm{yr}^{-1}}\right)^{1/2} \left(\frac{v_w}{2\times10^{3}\,\mathrm{km} \, \mathrm{s}^{-1}}\right)^{1/2} \\ \left(\frac{n_0}{5\times10^{3}\, \mathrm{cm}^{-3}}\right)^{-1/2} \left(\frac{v_*}{30\,\mathrm{km}\,\mathrm{s}^{-1}}\right)^{-1}\,\mathrm{pc}.
\label{eq:stagr}
\end{multline}

Given the parameters assumed above for the values of $\dot{M}$, $v_w$, $n_0$, and $v_*$, the stagnation radius is $\sim0.04$ pc, which is roughly consistent the observed distance between the star and the location nearest to it on the shell ($\sim0.07$ pc). Since the thickness of the dust-emitting region is unresolved, it is difficult to constrain the separation distance between the star and the shell to $\pm0.03$ pc. 

Interestingly, if region A is indeed a bowshock and not a pressure-driven blister then previous estimates of its age that are based on its size and the ionization front velocity for a blister HII region are not applicable. This is because the size of the region, or its stagnation radius, is only dependent on the stellar mass-loss rate, wind velocity, the star's velocity relative to the molecular cloud, and the density of the molecular cloud (Eq.~\ref{eq:stagr}). We derive a lower limit for the age of the region A heating source based on its velocity ($\sim30$ km/s; Yusef-Zadeh et al. 2010) and its apparent displacement from the western edge of the dust emission ($\sim0.3$ pc), which is assumed to be consistent with the edge of the molecular cloud. This calculation suggests a lower age limit of $t_\mathrm{age}\gtrsim 10^4$ yrs.

\subsection{The Western ``Ridges"}

The decreasing 31/37 color temperatures of the ridges from east to west suggest they are heated by a source in the vicinity of region A. Assuming the dust in the ridges is composed of 0.1 $\mu$m-sized silicate grains and heated by the stellar radiation from the region A heating source at a distance of $\sqrt{2}$ times the projected distance, Eq.~\ref{eq:Deq} can be used to predict equilibrium-heated dust temperatures of 60 and 47 K for ridges 2 and 3, respectively. The close agreement of the predicted dust temperatures with the color temperatures is strongly indicative that the ridges are heated by the same source. Interestingly, this excludes heating by the hot, young stars in the central cluster around Sgr A*, which is located $\sim140''$ to the southwest of ridge 3. At this distance with no intervening extinction, the central cluster would dominate the dust heating of ridge 3 and contribute significantly to heating dust in ridge 2. The additional heating from the central cluster ($L_\mathrm{cent}\sim2\times10^7$ $L_\odot$) would drive the dust temperatures of ridges 2 and 3 up to 65 K and 60 K, respectively. Since the difference between the observed 31/37 color temperatures of ridges 2 and 3 is much greater than the difference between these predicted dust temperatures it does not appear likely that the central cluster contributes significantly to heating the dust in the ridges. This is possibly due to the blockage of the UV light from this cluster by the intervening presence of the dense circumnuclear disk (Latvakoski et al. 1999; Lau et al. 2013), or by the dense ridge of molecular gas in the 50 km/s cloud lying to the west of the HII regions.

Position-velocity maps of the ridges from Yusef-Zadeh et al. (2010) show that ridges 1 and 2, which are the only ridges in the slit, exhibit radial velocities of $\sim50$ km/s, identical to the median velocity of the molecular cloud. Given their similar velocities to the molecular cloud and the increasing differential extinction from east to west, the ridges are likely elongated density enhancements on the surface of the 50 km/s molecular cloud heated and ionized by the radiation field of the region A heating source. 

\subsection{Heating Source Candidates}

\subsubsection{Regions A - C}

\begin{figure*}[h]
	\centerline{\includegraphics[scale=.40]{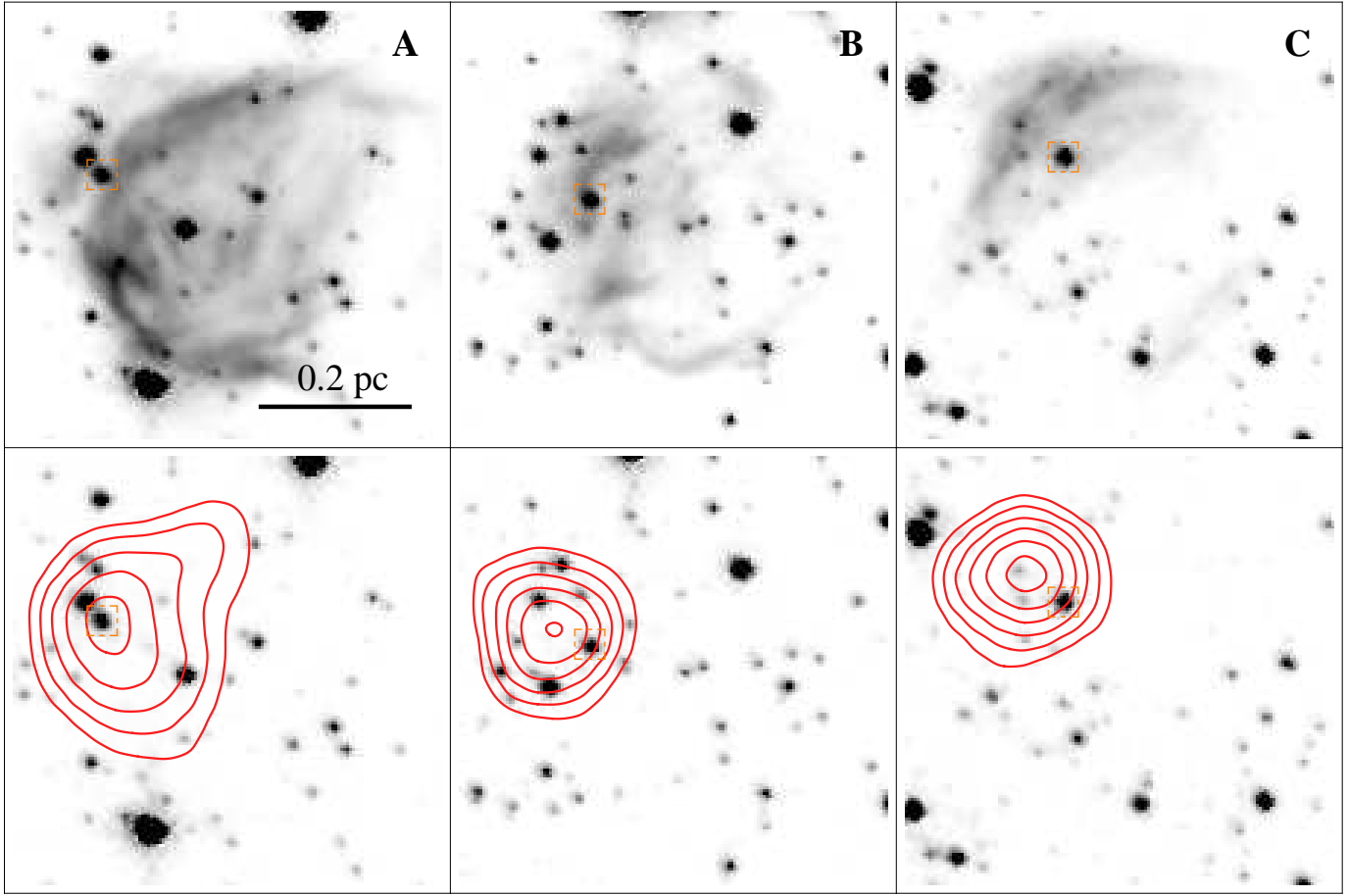}}
	\caption{(top row) Paschen-$\alpha$ (1.87 $\mu$m; Wang et al. 2010, Dong et al. 2011)) continuum image of regions A, B, and C with the heating source candidate marked in the orange, dotted squares. (bottom row) 1.90 $\mu$m image of the same fields as the top row with the same heating source candidates marked overlaid with the 19/37 color temperature contours. Color temperature contours correspond to 110, 115, 120, 125, and 130 K for A, 110, 115, 120, 125, 130, and 135 K for B, and 120, 125, 130, 135, 140, and 145 K for C.}
	\label{fig:GCHIIStars}
\end{figure*}

The 1.87 and 1.90 $\mu$m images reveal that numerous point-sources appear coincident along the line of sight within each of the regions A - C (Fig.~\ref{fig:GCHIIStars}a), which imposes difficulties in trying to determine the single heating source of the regions. Estimates of the projected location and total luminosity of the heating sources from the dust SED models provide possible candidates in the 1.87 and 1.90 $\mu$m fields (Tab.~\ref{tab:GCHIIStars}). The stars enclosed in the orange, dotted square in Fig.~\ref{fig:GCHIIStars}a and b exhibit a 1.9 $\mu$m flux consistent within $30\%$ of the 1.90 $\mu$m flux predicted by the heating sources of the DustEM SED models (assuming an O6 stellar atmosphere for the region A and B heating source and an O5 atmosphere for the C heating source). These heating source candidates are also suggestively located nearby the observed 19/37 color temperature peak. 

In the case of region A, the candidate lies only $\sim0.05$ pc away in projection from the approximated location of the heating source, which falls within the error annuli in Fig.~\ref{fig:GCAPlots}. The source $\sim0.2$ pc to the southwest of the candidate for A may also be a viable heating source given its eastern displacement from the center of curvature; however, its flux at 1.90 $\mu$m is greater than twice the predicted value, which would imply dust temperatures $\sim15$ K greater than the temperature determined for region A ($T_d\sim100$ K). It is interesting that the region A candidate is the same source identified by Mills et al. (2011), who claim it may be linked to the ``dark lane" structure at the northeastern edge that appears to separate the main shell of region A from a protrusion of emission at the northeast in Paschen-$\alpha$. Radio images of the region exhibit an identical structure to that seen in Paschen-$\alpha$ (Mills et al. 2011). 

\subsubsection{Region D}

A prominent point source appears coincident with the center of the western lobe of region D at 1.87 and 1.90 $\mu$m. The nature of this point source and its relation to region D has been disputed: Yusef-Zadeh et al. (2010) claim it is scattered light from a clump of gas and dust, and Mills et al. (2011) argue that it is actually the emission from the young stellar object that heats the region. Mills et al. (2011) claim that the source is unlikely a clump of dust and gas since that would imply enhanced emission from the ionized gas, which is not observed at Paschen-$\alpha$ or radio wavelengths. The 1.87/1.90 flux ratio also suggests the source has a purely stellar continuum (Mills et al. 2011). Our interpretation of this point source disagrees with both the claims of it being a gas and dust clump or the young stellar object based on its observed 1.90 $\mu$m flux. The 1.90 $\mu$m flux from this source is $\sim40$ times greater than the 1.90 $\mu$m flux predicted by the model heating source fitted to the region D SED, in which a stellar atmosphere of an O9 star (Mills et al. 2011) and an interstellar extinction value of $A_{P\alpha}=5.8$ (Mills et al. 2011) is assumed. Even if the amount of extinction towards the source was similar to that of regions A - C ($A_{P\alpha}=3.7$), the observed 1.90 $\mu$m flux would still be almost an order of magnitude greater than predicted. The point source is therefore most likely a star that is coincident with region D along the line of sight.

Based on the heating source luminosity from the SED model and an interstellar extinction value of $A_{P\alpha}=5.8$, the source should have a dereddened 1.90 $\mu$m flux of $\sim0.02$ Jy, whereas the bright point source has a dereddened 1.90 $\mu$m flux of $\sim0.11$ Jy. There is no $0.02$ Jy source detected in the vicinity of region D at 1.90 $\mu$m; however, the emission may be confused with the point spread function of the bright point source. 

\subsection{Region C Heating Source Stellar Type Inconsistency}

The inconsistency of the stellar classification of the heating source of region C between our model fits to the dust SEDs and the Lyman-continuum flux derived by Mills et al. (2011) presents an interesting issue. Mills et al. (2011) determine an O8.5 star while we require an O5 star, which implies a difference of almost an order of magnitude in stellar luminosity. Dust temperatures we observe are too high compared to the temperatures predicted by heating from the later-type O stars: an O8.5 star will heat 0.1 $\mu$m-sized silicate-type dust at distance of 0.2 pc away to a temperature of $\sim70$ K, which is significantly below the observed temperature of $\sim115$ K. Additionally, the total IR luminosity from region C, whose ``open" shell-like morphology suggests a small coverage factor, is similar to the total luminosity of an O8.5 star.
As shown in Sec.~4.1, the high observed dust temperatures cannot be explained by trapped Lyman-$\alpha$ heating. The heating contribution from the stellar winds is insignificant as well since 

\begin{multline}
\left(\frac{dE}{dt}\right)_\mathrm{KE}\sim 330 \left(\frac{\dot{M}}{10^{-6}\,M_\odot \, \mathrm{yr}^{-1}}\right) \\ \left(\frac{v_w}{2\times10^{3}\,\mathrm{km} \, \mathrm{s}^{-1}}\right)^{2} L_\odot << L_*.
\label{eq:wind}
\end{multline}

We propose that multiple stars are located in the vicinity of region C that are heating the dust and ionizing the gas. In order to reproduce both the observed dust temperatures and Lyman-continuum flux we require $\sim10$ O9.5 stars at a distance of 0.2 pc from the dust and gas. We were not able to reproduce the dust temperatures and Lyman-continuum flux with any other combination of O-type stars.

\section{Summary}
We have presented and studied images of the warm dust emissions from the Sgr A East HII region complex G-0.02-0.07 at 19.7, 25.2, 31.5, and 37.1 $\mu$m and arrive at the following conclusions:

- The general dust morphology of the three resolved HII regions (A, B, and C) agrees with that of the ionized gas as detected at Paschen-$\alpha$. However, in the case of region B, the ionized gas emission peaks several arcseconds interior to the dust emission peak at the northeast.

- Two faint IR sources, which are labeled as FIRS 1 and 2, are revealed at the FORCAST wavelengths. FIRS 1 is unresolved and has no ionized gas counterpart, and FIRS 2 is partially resolved (FWHM $\sim5''$ at 31.5 $\mu$m) and has a faint ionized gas counterpart observed in 6 cm emission (Zhao, Morris, \& Goss 2014). FIRS 1 appears coincident with the southern edge of a long linear filament, which is seen at 31.5 and 37.1 $\mu$m as well as Paschen-$\alpha$ and seemingly traces the surface of the 50 km/s molecular cloud (Mills et al. 2011). Model fits to the SED of FIRS 1 performed with DustEM show that it is heavily extinguished relative to the regions A, B, and C and indicate that the heating source has a total luminosity of $\sim1.7\times 10^4$ $L_\odot$ assuming that it is equivalent to the total, integrated IR luminosity. FIRS1 must therefore be deeply embedded since it is highly extinguished and has its heating source luminosity being intercepted and reradiated in the mid to far-IR. DustEM models fail to constrain the properties of FIRS 2.

- Two of the three Paschen-$\alpha$ ``ridges" to the west of region A as revealed by Mills et al. (2011) are detected and resolved at 31.5 and 37.1 $\mu$m. Based on the Paschen-$\alpha$ and 6 cm emission, there is a gradient of decreasing extinction west of ridge 1; however, the extinction estimates towards ridge 3 should be treated as an upper limit since there is non-thermal radio emission originating from the Sgr A East supernova remnant along this line of sight. Given the extinction gradient as well as a gradient of decreasing 31/37 color temperature from east to west, the ridges are interpreted as elongated density enhancements on the far surface of the 50 km/s molecular cloud heated and ionized by the region A heating source.

- DustEM model fits to the SEDs of regions A - D and FIRS 1 reveal the luminosity of the heating sources as well as the mass of the emitting dust (Tab.~\ref{tab:GCHIIProp}). For regions A - C, the spectral type of the heating sources we derive are slightly earlier than those estimated by Mills et al. (2011) from the Lyman-continuum flux. In the case of region C, where the inconsistency is most pronounced, we propose that there are multiple late-type O-stars heating the dust and ionizing the gas. The independently derived gas (Mills et al. 2011) and dust masses suggest a gas-to-dust mass ratio of $\sim55$, which we treat as a lower-limit since the emission from the dust traces deeper into the molecular cloud than the ionized gas emission. The presence of PAHs are required for the DustEM model fits to the SEDs of regions A - C.

\begin{figure*}[h]
	\centerline{\includegraphics[scale=.40]{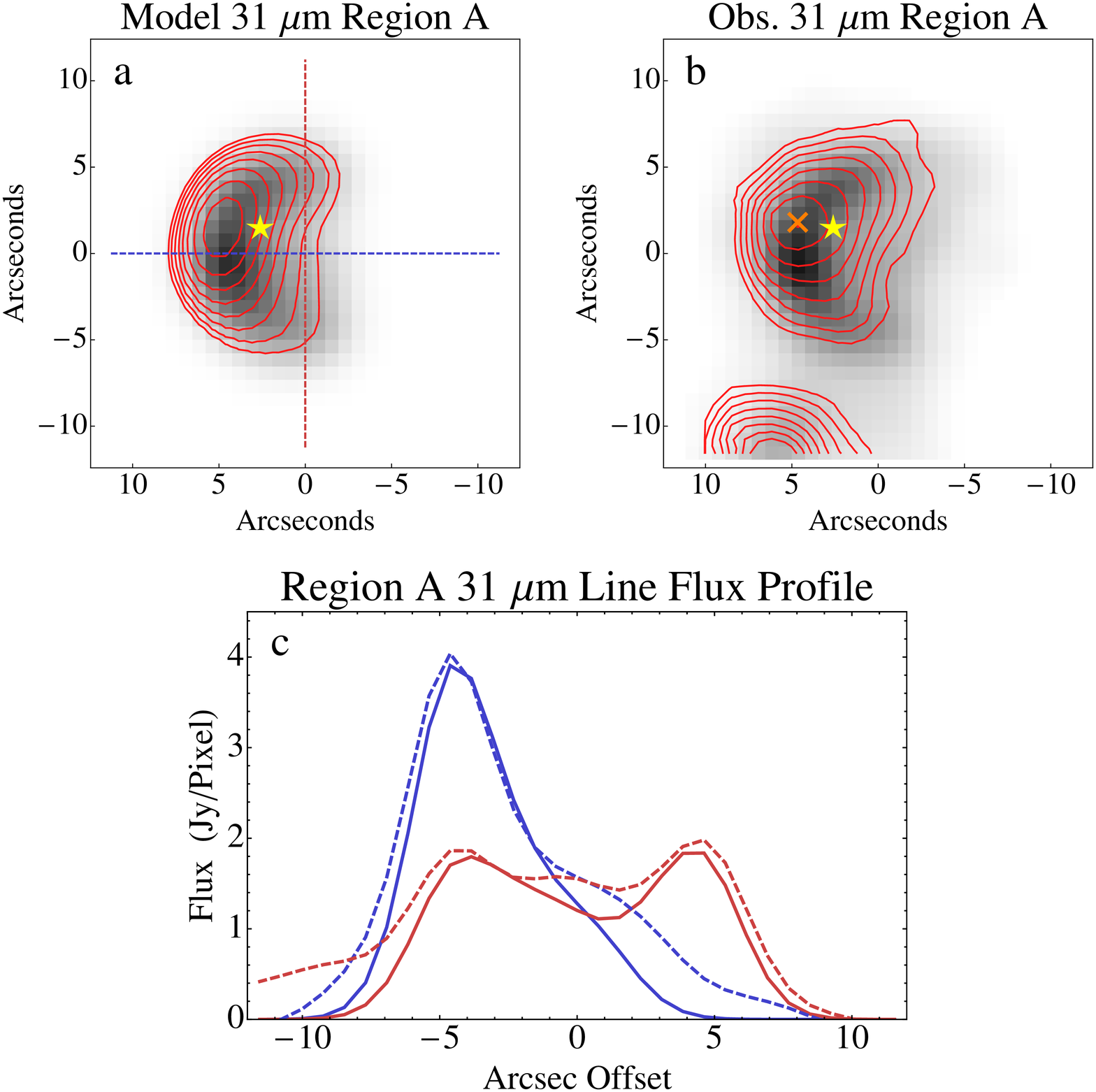}}
	\caption{(a) 31 $\mu$m hemispherical shell emission model of region A overlaid with the position of the heating source and 19/31 model color temperature contours with levels corresponding to 100, 105, 110, 115, 120, 125, 130, and 135 K.  (b) 31 $\mu$m observed flux map of region A overlaid with the position of the heating source and the observed 19/31 color temperature peak with contours corresponding to 100, 105, 110, 115, 120, 125, and 130 K. (c) Vertical (red) and horizontal (blue) flux cuts along the lines overlaid in (a) of both the observed (dashed) and model (solid) 31 $\mu$m emission.}
	\label{fig:GCAPlots}
\end{figure*}

- DustEM model fits to SEDs at north, east, and south locations in region A provide an estimate of the location of the heating source that is displaced northeast from the center of curvature (Fig.~\ref{fig:GCAPlots}). A hemispherical shell emission model of region A with the heating source at the estimated location agrees with the observed dust emission. The northeast displacement of the heating source and the kinematic properties of the gas of the region (Yusef-Zadeh et al. 2010) indicate that the region is a bow shock driven by stellar winds, as opposed to being an expanding blister formed by the overpressure of the ionized gas. Previous age estimates of the region that were made assuming the expansion was pressure driven are therefore not applicable. Using the velocity of the heating source and its displacement from the western edge of the dust emission, we estimate a lower limit of the heating source age to be $\sim10^4$ yrs.

- Candidate heating sources are identified for regions A - C in the Paschen-$\alpha$ continuum and 1.90 $\mu$m maps of the region that agree with the estimated flux of the heating sources as determined by the DustEM models. The positions of the candidate heating sources are also consistent with the locations of the 19/37 color temperature peaks.

- The bright point source coincident with the center of the western lobe of region D at 1.87 and 1.90 $\mu$m was claimed to be either the young stellar object heating and ionizing the gas and dust (Mills et al. 2011) or scattered light from a clump of gas and dust (Yusef-Zadeh et al. 2010). The source appears stellar in nature; however, the observed 1.90 $\mu$m flux is $\sim40$ times greater than the flux predicted for the heating source in the DustEM SED model. We therefore interpret this bright point source as a star coincident along the line of sight with region D.

\emph{Acknowledgments}. We would like to thank the rest of the FORCAST team, George Gull, Justin Schoenwald, and Chuck Henderson, the USRA Science and Mission Ops teams, and the entire SOFIA staff. We would also like to thank Betsy Mills for her valuable comments. This work is based on observations made with the NASA/DLR Stratospheric Observatory for Infrared Astronomy (SOFIA). SOFIA science mission operations are conducted jointly by the Universities Space Research Association, Inc. (USRA), under NASA contract NAS2-97001, and the Deutsches SOFIA Institut (DSI) under DLR contract 50 OK 0901. Financial support for FORCAST was provided by NASA through award 8500-98-014 issued by USRA.

\clearpage

\begin{deluxetable*}{c|ccccccccc}
\tablecaption{Observed fluxes of the Sgr A East HII Regions and Faint IR Sources in Jy}
\tablewidth{0pt}
\tablehead{ Region &$F_{3.6}$\tablenotemark{a} & $F_{4.5}$\tablenotemark{a} & $F_{5.8}$\tablenotemark{a} & $F_{8.0}$\tablenotemark{a} & $F_{12.8}$\tablenotemark{b} & $F_{19.7}$ & $F_{25.2}$ & $F_{31.5}$ & $F_{37.1}$ }

\startdata
	A& 0.28 & 0.40 & 1.85 &  5.80 & & 119 & 271 & 354 & 389\\ 
	B& 0.18 & 0.39 & 1.16 &  3.14 &  & 35.6 & 80.5 & 106 & 116\\ 
	C& 0.15 & 0.26 & 1.05 &  3.76 & & 57.7 & 110 & 120 & 119\\ 
	D& 0.19 & 0.75 & 2.22 &  4.47 & 7.50 & 11.7 & 32.5 & 35.3 & 28.9 \\ 
	FIRS 1& $\lesssim$0.0048& 0.055 & 0.29 & 0.48 &  & 0.63 & 2.45 & 3.55 & 4.00 \\ 
	FIRS 2& &  & 0.04 & 0.10  &  & 0.42 & 2.29 & 3.02 & 3.35 \\ 
\enddata
\tablenotetext{a}{Fluxes from Spitzer/IRAC (Stolovy et al. 2006)}
\tablenotetext{b}{Flux from IRTF/TEXES (Yusef-Zadeh et al. 2010)}
	\label{tab:GCHIIFlux}
\end{deluxetable*}

\begin{deluxetable*}{c|cccccccc}
\tablecaption{Summary of the Sgr A East HII Complex and FIRS 1 Properties}
\tablewidth{0pt}
\tablehead{ Region& $d$ (pc)&$T_\mathrm{d}$\tablenotemark{a} (K) & $L_\mathrm{*}$\tablenotemark{a} ($10^4 \, L_\odot$) & $L_\mathrm{IR}/L_\mathrm{*}$ & $M_\mathrm{dust}$\tablenotemark{a} ($M_\odot$) & $\frac{M_\mathrm{HII}}{M_\mathrm{dust}}$\tablenotemark{b} & $\frac{Y_\mathrm{VSG}}{Y_\mathrm{LG}}$\tablenotemark{a}& $\frac{Y_\mathrm{PAH}}{Y_\mathrm{VSG}}$\tablenotemark{a} }

\startdata
	A& 0.45 & 98 & 22 & 0.77 & 0.039 & 56 & 0.009 & 1.2 \\ 
	B& 0.34 & 96 & 19 & 0.28 & 0.013 & 60 & 0.038 & 0.39\\ 
	C&  & 111 & 40 & 0.17 & 0.009 & 45 & 0.016 & 0.79\\ 
	D& 0.16 (dec) & 131 & 4.7& 1.0 & 0.002 & 65 & 0.37 & $\lesssim0.3$ \\ 
	FIRS 1& & 163 & 1.7 & 1.0 & 0.0002 &  & 0.033 & $\lesssim0.02$ \\ 
\enddata
\tablecomments{$d$ is the observed diameter of the region, $T_d$ is the the volume-average dust temperature, $L_\mathrm{*}$ is the heating source luminosity, $L_\mathrm{IR}/L_\mathrm{*}$ is the integrated IR to heating source luminosity ratio, $M_\mathrm{dust}$ is the dust mass, $\frac{M_\mathrm{HII}}{M_\mathrm{dust}}$ is the gas-to-dust ratio, $\frac{Y_\mathrm{VSG}}{Y_\mathrm{LG}}$ is the VSG-to-LG abundance ratio, and $\frac{Y_\mathrm{PAH}}{Y_\mathrm{VSG}}$ is the PAH-to-VSG abundance ratio.}
\tablenotetext{a}{Determined by DustEM model fit to the SEDs}
\tablenotetext{b}{$M_\mathrm{HII}$ derived from free-free/Paschen-$\alpha$ emission (Mills et al. 2011)}
	\label{tab:GCHIIProp}
\end{deluxetable*}

\begin{deluxetable*}{c|ccccc}
\tablecaption{Dust Properties of the Region A ``Ridges"}
\tablewidth{0pt}
\tablehead{ Ridge& $d_\mathrm{proj}$ (pc)&  $T_{31/37}$ (K) & $T_{eq}$ (K)& $\tau_{37}$ & $A_{P_\alpha}$ }

\startdata
	1& 0.4 &  &  &  & $3.1\pm0.2$\\ 
	2& 0.5 &  $64\pm4$ & 60 & $\sim0.01$ &  $3.7\pm0.2$\\ 
	3& 1.0   & $40\pm4$ & 47 & $\gtrsim0.04$ & $\lesssim4.5$\\ 
\enddata

\tablecomments{$d_\mathrm{proj}$ is projected distance of the ridges from region A, $T_{31/37}$ is the observed 31/37 color temperature, $T_{eq}$ is the predicted dust temperature assuming a distance of $d_\mathrm{proj}$ multiplied by a $\sqrt{2}$ projection factor, $\tau_{37}$ is the 37.1 $\mu$m optical depth, and $A_{P_\alpha}$ is the extinction magnitude at 1.87 $\mu$m.}
	\label{tab:GCHIIA}
\end{deluxetable*}

\begin{deluxetable*}{ccccccc}
\tablecaption{Region A Dust Emission Model Parameters}
\tablewidth{0pt}
\tablehead{ $r_0$ (pc) &$\Delta t$ (pc) & $n_S$ ($\mathrm{cm}^{-3}$)& $L_*$ ($L_\odot$) &$\Delta x$ &$\Delta y$ & $a$ ($\AA$) }

\startdata
	0.18& 0.06 & 2100  & $2\times10^5$ &  -3'' & +1'' & 100\\ 
\enddata

\tablecomments{$r_0$ is the inner radius, $\Delta t$ is the thickness of the shell, $n_S$ is the gas density in the southern half of the shell, $L_*$ is the stellar luminosity of the heating source, $\Delta x$ and $\Delta y$ are the RA and dec displacement of the heating source with respect to the center of the shell, and $a$ is the size of the silicate grains in the shell.}
	\label{tab:GCHIIAMod}
\end{deluxetable*}

\begin{deluxetable*}{c|cccc}
\tablecaption{Dereddened Candidate Heating Source and Model 1.90 $\mu$m Fluxes}
\tablewidth{0pt}
\tablehead{ & A & B & C & D\tablenotemark{a}}

\startdata
	Candidate $F_{1.90}$ (Jy)& 0.085 & 0.055  & 0.101 & 0.108\\ 
	Model $F_{1.90}$ (Jy)& 0.063 & 0.054 & 0.09 & 0.019 \\ 
	$A_{P\alpha}$& 3.7 & 3.7 & 3.7 & 5.8\\ 
\enddata

\tablenotetext{a}{The candidate flux is that of the heating source proposed by Mills et al. (2011)}
	\label{tab:GCHIIStars}
\end{deluxetable*}

\newpage

\vfill



\begin{thebibliography}{99}


\bibitem[An et al.(2011)]{2011ApJ...736..133A} An, D., Ram{\'{\i}}rez, 
S.~V., Sellgren, K., et al.\ 2011, \apj, 736, 133 
\bibitem[Cardelli et al.(1989)]{1989ApJ...345..245C} Cardelli, J.~A., Clayton, G.~C., \& Mathis, J.~S.\ 1989, \apj, 345, 245 
\bibitem[Compi{\`e}gne et 
al.(2011)]{2011A&A...525A.103C} Compi{\`e}gne, M., Verstraete, L., Jones, A., et al.\ 2011, \aap, 525, A103 
\bibitem[Desert et 
al.(1986)]{1986A&A...160..295D} Desert, F.~X., Boulanger, F., \& Shore, S.~N.\ 1986, \aap, 160, 295 
\bibitem[Desert et 
al.(1990)]{1990A&A...237..215D} Desert, F.-X., Boulanger, F., \& Puget, J.~L.\ 1990, \aap, 237, 215 
\bibitem[Dong et al.(2011)]{2011MNRAS.417..114D} Dong, H., Wang, Q.~D., Cotera, A., et al.\ 2011, \mnras, 417, 114 
\bibitem[Dong et al.(2012)]{2012MNRAS.425..884D} Dong, H., Wang, Q.~D., 
\& Morris, M.~R.\ 2012, \mnras, 425, 884
\bibitem[Draine 
\& Li(2001)]{2001ApJ...551..807D} Draine, B.~T., \& Li, A.\ 2001, \apj, 551, 807 
\bibitem[Draine(2011)]{2011piim.book.....D} Draine, B.~T.\ 2011, Physics of the Interstellar and Intergalactic Medium by Bruce T.~Draine.~Princeton University Press, 2011.~ISBN: 978-0-691-12214-4
\bibitem[Ekers et 
al.(1983)]{1983A&A...122..143E} Ekers, R.~D., van Gorkom, J.~H., Schwarz, U.~J., \& Goss, W.~M.\ 1983, \aap, 122, 143 
\bibitem[Fritz et al.(2011)]{2011ApJ...737...73F} Fritz, T.~K., Gillessen, S., Dodds-Eden, K., et al.\ 2011, \apj, 737, 73 
\bibitem[Goss et al.(1985)]{1985MNRAS.215P..69G} Goss, W.~M., Schwarz, 
U.~J., van Gorkom, J.~H., \& Ekers, R.~D.\ 1985, \mnras, 215, 69P 
\bibitem[Herter et al.(2012)]{2012ApJ...749L..18H} Herter, T.~L., Adams, J.~D., De Buizer, J.~M., et al.\ 2012, \apjl, 749, L18
\bibitem[Herter et al.(2013)]{2013PASP..125.1393H} Herter, T.~L., Vacca, 
W.~D., Adams, J.~D., et al.\ 2013, \pasp, 125, 1393 
\bibitem[Krabbe et al.(1991)]{1991ApJ...382L..19K} Krabbe, A., Genzel, R., 
Drapatz, S., \& Rotaciuc, V.\ 1991, \apjl, 382, L19 
\bibitem[Latvakoski et al.(1999)]{1999ApJ...511..761L} Latvakoski, H.~M., 
Stacey, G.~J., Gull, G.~E., \& Hayward, T.~L.\ 1999, \apj, 511, 761 
\bibitem[Lau et al.(2013)]{2013ApJ...775...37L} Lau, R.~M., Herter, T.~L., 
Morris, M.~R., Becklin, E.~E., \& Adams, J.~D.\ 2013, \apj, 775, 37
\bibitem[Mathis et al.(1977)]{1977ApJ...217..425M} Mathis, J.~S., Rumpl, 
W., \& Nordsieck, K.~H.\ 1977, \apj, 217, 425 
\bibitem[Mauerhan et al.(2010)]{2010ApJ...725..188M} Mauerhan, J.~C., 
Cotera, A., Dong, H., et al.\ 2010, \apj, 725, 188 
\bibitem[Mills et al.(2011)]{2011ApJ...735...84M} Mills, E., Morris, M.~R., 
Lang, C.~C., et al.\ 2011, \apj, 735, 84 
\bibitem[Morris \& Serabyn(1996)]{1996ARA&A..34..645M} Morris, M., \& Serabyn, E.\ 1996, \araa, 34, 645
\bibitem[Muno et al.(2009)]{2009ApJS..181..110M} Muno, M.~P., Bauer, F.~E., 
Baganoff, F.~K., et al.\ 2009, \apjs, 181, 110 
\bibitem[Nagata et al.(1995)]{1995AJ....109.1676N} Nagata, T., Woodward, 
C.~E., Shure, M., \& Kobayashi, N.\ 1995, \aj, 109, 1676 
\bibitem[Okuda et al.(1990)]{1990ApJ...351...89O} Okuda, H., Shibai, H., 
Nakagawa, T., et al.\ 1990, \apj, 351, 89 
\bibitem[Osterbrock 
\& Ferland(2006)]{2006agna.book.....O} Osterbrock, D.~E., \& Ferland, G.~J.\ 2006, Astrophysics of gaseous nebulae and active galactic nuclei, 2nd.~ed.~by D.E.~Osterbrock and G.J.~Ferland.~Sausalito, CA: University Science Books, 2006,
\bibitem[Reid(1993)]{1993ARA&A..31..345R} Reid, M.~J.\ 1993, \araa, 31, 345
\bibitem[Serabyn et al.(1992)]{1992ApJ...395..166S} Serabyn, E., Lacy, 
J.~H., \& Achtermann, J.~M.\ 1992, \apj, 395, 166 
\bibitem[Sjouwerman et 
al.(2002)]{2002A&A...391..967S} Sjouwerman, L.~O., Lindqvist, M., van Langevelde, H.~J., \& Diamond, P.~J.\ 2002, \aap, 391, 967 
\bibitem[Shull(1980)]{1980ApJ...238..860S} Shull, J.~M.\ 1980, \apj, 238, 
860 
\bibitem[Stolovy et al.(2006)]{2006JPhCS..54..176S} Stolovy, S., Ramirez, 
S., Arendt, R.~G., et al.\ 2006, Journal of Physics Conference Series, 54, 
176 
\bibitem[Wang et al.(2010)]{2010MNRAS.402..895W} Wang, Q.~D., Dong, H., Cotera, A., et al.\ 2010, \mnras, 402, 895 
\bibitem[Yusef-Zadeh \& Morris(1987)]{1987ApJ...320..545Y} Yusef-Zadeh, F., \& Morris, M.\ 1987, \apj, 320, 545
\bibitem[Yusef-Zadeh et al.(2010)]{2010ApJ...725.1429Y} Yusef-Zadeh, F., 
Lacy, J.~H., Wardle, M., et al.\ 2010, \apj, 725, 1429 
\bibitem[Zhao, Morris, \& Goss(2014]{z} Zhao, J.-H., Morris, M.~R., \& Goss, W.~M. 2014, in preparation
\end{thebibliography}
\end{document}